\documentclass[11pt]{article}
\usepackage{amsmath,amssymb, diagrams}
\usepackage{epsfig}
\newcommand{\be}{\begin{equation}}
\newcommand{\ee}{\end{equation}}
\newcommand{\bea}{\begin{array}}
\newcommand{\ea}{\end{array}}
\newcommand{\beqa}{\begin{eqnarray}}
\newcommand{\eeqa}{\end{eqnarray}}
\newcommand{\nn}{\nonumber}
\begin{document}

\begin{flushright}
SU-4252-823 \\
SINP/TNP/06-02 \\
IMSC/2006/3/4 \\
DIAS-STP-06-03 \\
\end{flushright}

\begin{flushright}   
\today
\end{flushright} 
\begin{center}
\vskip 2em
{\LARGE Noncommutative Two Dimensional Gravities} \\
\vskip 2em
\centerline{A.~P.~Balachandran$^{a}$,
T.~R.~Govindarajan$^{b}\footnote{On leave of absence from Institute 
for Mathematical Sciences,  C. I. T. Campus Taramani,Chennai 600 113,
India, E-mail:trg@imsc.res.in.} \footnote{At Max Planck Institute for
Gravitation Physics, AEI, Golm, Germany, after $25^{th}$ 
February until $30^{th}$ June.}$, Kumar~S.~Gupta$^{c}$,
Se\c{c}kin~K\"{u}rk\c{c}\"{u}o\v{g}lu$^{d}$}

\vskip 2em

\centerline{\sl  $^a$  Department of Physics}
\centerline{\sl  Syracuse University}
\centerline{\sl  Syracuse NY 13244-1130 USA}
{\sl  e-mails:} \hskip 2mm bal@phy.syr.edu

\vskip 1em

\centerline{\sl $^b$ Instituto de F\'{\i}sica}
\centerline{\sl Universidade de S\~{a}o Paulo} 
\centerline{\sl C.P. 66318, S\~{a}o Paulo, SP, 05315-970, Brazil}
{\sl e-mails:}  \hskip 2mm tiargi@fma.if.usp.br

\vskip 1em

\centerline{\sl $^c$ Saha Institute of Nuclear Physics}
\centerline{\sl Theory Division, 1/AF Bidhannagar}
\centerline{\sl  Kolkata 700064, India.}
{\sl  e-mails:} \hskip 2mm kumars.gupta@saha.ac.in

\vskip 1em

\centerline{\sl  $^d$  School of Theoretical Physics}
\centerline{\sl  Dublin Institute for Advanced Studies}
\centerline{\sl  10 Burlington Road}
\centerline{\sl  Dublin 4, Ireland}
{\sl  e-mails:}  \hskip 2mm seckin@stp.dias.ie
\end{center}

\vskip 2em

\begin{abstract}

We give formulations of noncommutative two dimensional gravities in terms of 
noncommutative gauge theories. We survey their classical solutions and
show that solutions of the corresponding commutative theories continue to 
be solutions in the noncommutative theories as well. We argue that the 
existence of ``twisted'' diffeomorphisms, recently introduced in \cite{wess}, 
is crucial for this conclusion. 
 
\end{abstract}

\vskip 1em

{\it Dedicated to Rafael Sorkin, our friend and teacher, and a true and creative seeker of knowledge.}

\newpage

\setcounter{footnote}{0}
 
\section{Introduction}

Recently progress have been made in formulating gravity theories on
noncommutative(NC) spaces. In \cite{wess}, a new approach has been
developed to restore the action of diffeomorphisms as a symmetry of the 
Groenewold-Moyal type NC spacetimes. . Using this new approach, the authors
of \cite{wess} were able to construct a noncommutative version of the 
Einstein-Hilbert action, which is invariant under the deformed or 
``twisted'' diffeomorphisms. However, due to the complicated nature of the
resulting expressions, it is not easy to extract the physical
predictions of the new theory [However, for recent progress along this 
line see \cite{wess1}]. Thus it could be practically advantageous to study 
the consequences of these developments in the less complicated setting 
of two dimensional gravities. The latter are well-known in the literature and 
elegantly formulated as gauge theories \cite{Jackiw, Isler, Verlinde, Mandal, 
Witten, Callan, Cangemi1}.

In this article, we construct ``twisted'' generally covariant
noncommutative gauge theories describing noncommutative gravity 
models on two dimensional noncommutative spaces. We analyze their 
classical solutions and show that they are the same as the solutions 
of their corresponding commutative theories. We argue that the presence 
of ``twisted'' diffeomorphisms is essential for the latter conclusion.

Our results for the case of the noncommutative version of the standard
two dimensional gravity theory bear strong similarities with the
earlier work of Cacciatori et. al. \cite{Klemm}, our difference being in the
class of diffemorphisms the relevant action admits and its implementation. 

Our results in this article are also consistent with the results of 
work in progress of Balachandran et. al. \cite{Balgravity}, where
another approach is being developed to understand the action of 
``twisted'' symmetries. It has as one of its predictions that 
gravity theories without sources on the Groenewold-Moyal plane
have the solutions of the corresponding commutative theories.

The need for ``twisted'' symmetries in formulating NC two dimensional gravities
have also been recently argued in \cite{Vassilevich}. 

This paper is organized as follows. In section 2 we discuss NC gauge theory formulation of 
two dimensional gravity with zero cosmological constant and show that it has trivial solutions. 
In section 3 we formulate the NC gauge theories for two dimensional gravities with nonzero cosmological
constant. Here we give the appropriate action and demonstrate its invariances under NC gauge transformations and 
``twisted'' diffeomorphisms. In section 4, we study the classical solutions of the theories formulated in 
section 3 and arrive at the result that they are same as that of their corresponding commutative theories. 
 
\section{Noncommutative Gauge Theory for Two Dimensional Gravity with Zero Cosmological Constant}

Let us formulate two  dimensional noncommutative gravity using the 
noncommutative version of the SO(1,1) gauge group.
In the commutative case, we can think of the $SO(1,1)$ gauge group as
generated by a single Pauli matrix, say $\sigma_2$. It is
well-known that gravity based on this group is completely
trivial. We will see below that the same holds for gravity
based on the noncommutative version of the $SO(1,1)$ gauge group. In this case, the 
gauge group has two commuting generators, $\sigma_2$ and the $2 \times
2$ identity matrix $I$, the latter arising due to the noncommutativity of space-time.
  
Consider the connection one form
\begin{equation}
A = \omega \sigma_2 + f I
\end{equation}
where $\omega$ is the spin connection and $f$ is an additional one form.
The corresponding curvature two form is 
\beqa
F &=& dA + A \wedge_* A \nonumber \\
  &=& (d\omega + f \wedge_*\omega + \omega \wedge_*f)~\sigma_2 + (df + \omega \wedge_* \omega
 + f \wedge_* f)~I \nonumber \\
 & \equiv & F^1 \sigma_2 + F^2 I \, ,
\label{eq:two}
\eeqa
where $\wedge_*$ is understood to be the ordinary wedge product, except 
that the components of differential forms are now being multiplied with
the Groenewold-Moyal $*$-product\footnote{For two functions $f\,, g \in {\cal A}_\theta({\mathbb R}^2)$ 
the Groenewold-Moyal $*$-product is defined as:
\begin{equation}
f*g(x)~=~f(x) e^{\frac{i}{2} \overleftarrow{\partial}_\mu 
\theta^{\mu\nu}\overrightarrow{\partial}_\nu}g(x) \,, \nonumber \
\end{equation}
where $\theta^{\mu \nu} = \theta \varepsilon^{\mu \nu}, \, (\varepsilon^{01} =1) $ and $\theta$
is the noncommutativity parameter.}.

Let us also introduce a two component scalar field $\phi = (\phi_1 \sigma_2 + \phi_2 I)$. 
Using $\phi$ and $F$ we can form the gauge invariant action
\beqa
S &=& \frac{1}{2} \int Tr \, (\phi * F) \nonumber \\
&=& \frac{1}{2} \int \phi_1 * F^1 + \phi_2 * F^2 \,.
\label{eq:three}
\eeqa

$S$ is invariant under infinitesimal gauge transformations
\be
F\rightarrow F + i \lbrack \upsilon \,, F \rbrack_* \,, \quad \phi 
\rightarrow \phi + i \lbrack \upsilon \,, \phi \rbrack_* \,,
\ee
where $\upsilon = \upsilon_1 \sigma_2 + \upsilon_2 I$ is the gauge transformation parameter.

Using (\ref{eq:two}) and (\ref{eq:three}) and the fact that one $*$-product can be removed under the integral 
sign we see the nonlinear terms in the integrand of (\ref{eq:three}) either become zero or reduce to total derivatives.
Thus, we can rewrite $S$ as 
\beqa
S &=& \frac{1}{2} \int \phi_1 * d \omega + \phi_2 * d f \nonumber \\
&=& \frac{1}{2} \int \phi_1~d \omega + \phi_2 ~ d f \,.
\label{eq:ac2}
\eeqa
From (\ref{eq:ac2}) we infer that $d \omega = df = 0$. They give
trivial solutions for gravity.

It will become clear after the discussion in section 3 that the 
action (\ref{eq:ac2}) is indeed invariant under ``twisted'' 
diffemorphisms.

Hereafter we focus on theories with nonzero cosmological constant.

\section{Noncommutative Gauge Theories for Two Dimensional Gravity with Nonzero Cosmological Constant}

\subsection{Generalities}

Let us now direct our attention to the formulation of possible gauge
theories with nonzero cosmological constant and with or without a dilaton.
They are based on the gauge group $U(1,1) \approx SO(2,1) \times U(1)$ and 
its contractions. The presence of the extra $U(1)$ factor is due to the 
noncommutativity of the theory. 
 
The associated Lie algebra $so(2,1) \oplus u(1)$ is generated 
by $P_a, J$ and $I$ ($a = 0,1$). The commutation relations among 
these generators are given by
\begin{equation}
\lbrack P_a , P_b \rbrack = -\frac{1}{2} \frac{\Lambda}{s} 
\varepsilon_{ab} (2J - sI) \,, \quad  
\lbrack P_a \,, J \rbrack = {\varepsilon_a}^b P_b \,, \quad 
\lbrack P_a \,, I \rbrack = \lbrack J \,, I \rbrack = 0 \,, \quad (\varepsilon^{01} = 1) \,.
\label{eq:e1}
\end{equation}   
Thus $I$ is a central element. Here $\Lambda$ is the cosmological constant 
and $s$ is a dimensionless parameter whose role will be explained below. 
For the generators we will sometimes use the notation
\be
(Q_a, Q_2, Q_3) \equiv (P_a , J, I) \,,~~a \in \{0,1\}
\ee
in the text. 

For finite values of $\Lambda$ and $s$, the commutation relations in 
({\ref{eq:e1}) are those of the standard $so(2,1) \oplus u(1)$ Lie 
algebra as can be seen by making the substitutions $\frac{\Lambda}{s} 
\rightarrow \Lambda$ and $(J - \frac{1}{2} s I) \rightarrow J$. 
Keeping $\Lambda$ finite and letting $s \rightarrow \infty$ 
results in what is known as the centrally extended Poincar\'e algebra
\cite{Cangemi1}.  We also recall that the gauge theory 
formulation of the two dimensional gravity model is based on $so(2,1)$, while the centrally 
extended Poincar\'e algebra is required for the formulation ``string-inspired'' 
gravity \cite{Cangemi1}.

Throughout the paper we work with the fundamental representation 
of (\ref{eq:e1}). It is given by:
\begin{equation}
P_0 = \frac{1}{2} \sqrt{\frac{\Lambda}{s}} i \sigma_3 \,, \quad P_1 = 
\frac{1}{2} \sqrt{\frac{\Lambda}{s}} \sigma_1 \,, \quad J =
\frac{1}{2}(\sigma_2 + s I) \,,
\end{equation} 
where $\sigma_i\,, (i= 1,2,3)$ as usual denote the Pauli matrices.
In this representation the following relations hold:
\begin{gather}
J^2 = s J + \frac{1-s^2}{4} I \,, \quad  \lbrace J , P_a \rbrace = s
P_a \,, \quad  \lbrace P_a , P_b \rbrace = - \frac{\Lambda}{2 s} h_{ab} I \nonumber \\
Tr P_a P_b = -\frac{\Lambda}{2 s} h_{ab} \,, \quad Tr J^2 = \frac{1+s^2}{2} \,, \quad Tr
I = 2 \,,
\end{gather}
where $h_{ab} = diag(1, -1)$ and $\{.,.\}$ denote anticommutators. 

Let us consider the connection one form $A$. It is composed of the {\it
zweibein}'s~$e^a (a=0,1)$, the spin connection $\omega$ and the additional 
one form $k$. Expanding in the Lie algebra basis, $A$ reads    
\be
A := A^\alpha Q_\alpha = e^a P_a + \omega J + \frac{\Lambda}{2} k I
\,, \quad (\alpha = 0,1,2,3) \,.
\label{eq:connection}
\ee

We compute the curvature associated to $A$ in a straightforward fashion using  
\be
F = d A + A \wedge_* A \,.
\ee 
We find 
\begin{multline}
F = \Big \lbrack d e^a + \frac{1}{2} {\varepsilon_b}^a (e^b \wedge_* \omega
-\omega \wedge_* e^b) +\frac{1}{2} \Lambda ( k \wedge_* e^a
+ e^a \wedge_* k) + \frac{s}{2} (e^a \wedge_* \omega + \omega \wedge_*
e^a) \Big \rbrack P_a + \\
\Big \lbrack d \omega +  s \omega \wedge_* \omega  -\frac{\Lambda}{2s}
\varepsilon_{ab} e^a \wedge_* e^b + \frac{\Lambda}{2} 
(k \wedge_* \omega + \omega \wedge_* k) \Big \rbrack J + \\
\Big \lbrack \frac{\Lambda}{2} d k + \frac{\Lambda^2}{4} k \wedge_* k -
\frac{\Lambda}{4 s} h_{ab} e^a \wedge_* e^b   + 
\frac{\Lambda}{4} \varepsilon_{ab} e^a \wedge_* e^b +
\frac{1-s^2}{4} \omega \wedge_* \omega \Big \rbrack I  \,.
\end{multline}

Under the infinitesimal gauge transformations generated by $\upsilon = \upsilon^a P_a 
+ \upsilon^2 J + \upsilon^3 I$, we have
\beqa
A \longrightarrow A^\prime &=& A + i D^* \upsilon\,, \quad 
D^* \upsilon = d \upsilon + i \lbrack \upsilon \,, A \rbrack_* \nonumber \\ 
F \longrightarrow F^\prime &=& F + i \lbrack \upsilon \,, F \rbrack_* \,,
\label{eq:gauget}
\eeqa
while under finite gauge transformations
\beqa
A \longrightarrow A^\prime &=& e^{i \upsilon} * ( A + d ) * (e^{i \upsilon})^{-1}_* \,, \nn \\
F \longrightarrow F^\prime &=& e^{i \upsilon} * F * (e^{i \upsilon})^{-1}_* \,.
\eeqa
Note that in above $(e^{i \upsilon})^{-1}_*$ is the $*$-inverse of
$(e^{i \upsilon})$.

\subsection{The Action}

We now give the gauge theory action describing NC gravity theories with nonzero 
cosmological constant, in two dimensions. Generalizing from the commutative theory
we write,
\begin{equation}
S = \int Tr (\xi * F) \,,
\label{eq:action1}
\end{equation} 
Here we introduced the $4$-component scalar field 
\be
\xi = -\frac{2s}{\Lambda} \eta^a P_a +  \frac{2}{1+s^2} \eta^2 J + 
\frac{1}{\Lambda} \eta^3 I  \,,
\label{eq:scalarf1}
\ee
and the trace is taken over the Lie algebra basis. We note the peculiar 
factors in front of the component fields in (\ref{eq:scalarf1}); in the action
$S$, they cancel with the factors coming from the traces.   

In terms of the component fields, the action $S$ reads 
\begin{multline}
S = \int \eta_a * \Big \lbrack d e^a + \frac{1}{2}
{\varepsilon_b}^a (e^b \wedge_* \omega -\omega \wedge_* e^b) 
+\frac{\Lambda}{2} (k \wedge_* e^a + e^a \wedge_* k ) + \frac{s}{2} 
(e^a \wedge_* \omega + \omega \wedge_* e^a) \Big \rbrack + \\
\int \eta_2 * \Big \lbrack d \omega + s \omega \wedge_* \omega - 
\frac{\Lambda}{2s} \varepsilon_{ab} e^a \wedge_* e^b + \frac{\Lambda}{2} 
(k \wedge_* \omega + \omega \wedge_* k) \Big \rbrack + \\
\int \eta_3 * \Big \lbrack d k + \frac{\Lambda}{2} k \wedge_* k - 
\frac{1}{2} h_{ab} e^a \wedge_* e^b  + \frac{1}{2} \varepsilon_{ab}
e^a \wedge_* e^b + \frac{1-s^2}{2 \Lambda } w \wedge_* \omega  \Big \rbrack \,.
\label{eq:action2}
\end{multline}
Let us note that several terms in this action do in fact vanish. 
Recalling that one $*$-product can be removed under the integral 
and removing the $*$ in the $\wedge_*$, we find that the $3^{rd}$ and
$4^{th}$ terms of the first integral, the $2^{nd}$ and $4^{th}$ terms 
of the second integral and the $2^{nd}$ and $3^{rd}$ and $5^{th}$ terms of
the third integral are zero. Thus we have
\begin{multline}
S = \int \eta_a * \Big ( d e^a + \frac{1}{2}
{\varepsilon_b}^a (e^b \wedge_* \omega -\omega \wedge_* e^b) \Big ) + 
\eta_2 * \Big ( d \omega - \frac{\Lambda}{2s} 
\varepsilon_{ab} e^a \wedge_* e^b \Big ) + \\
\eta_3 * \Big ( d k + \frac{1}{2} \varepsilon_{ab}
e^a \wedge_* e^b \Big ) \,.
\label{eq:action3}
\end{multline}

$S$ is invariant under both the infinitesimal NC gauge transformations and
``twisted'' diffeomorphisms. We now explicitly demonstrate these invariances.

\subsection{Symmetries}

{\it Gauge Invariance:} \\

The action given in (\ref{eq:action1}) is gauge invariant under 
the infinitesimal gauge transformation given in (\ref{eq:gauget}) 
for the curvature $F$ and the standard infinitesimal transformation 
law of scalar fields in the NC gauge theories:
\be
\xi \rightarrow \xi + i \lbrack \upsilon \,, \xi \rbrack_*
\label{eq:adjoint_trans}
\ee 
Explicitly, we have 
\beqa
\lefteqn{S(\xi +\delta \xi , F +\delta F)} \nonumber \\ 
\quad \quad &=& \int Tr ( \xi + i \lbrack \upsilon \,, \xi \rbrack_*)
*( F + i \lbrack \upsilon \,, F \rbrack_*) \nonumber \\
\quad \quad &=& \int Tr \left ( \xi * F + i \xi * \lbrack \upsilon \,, F \rbrack_*
+ i \lbrack \upsilon \,, \xi \rbrack_* * F + {\cal O}(\upsilon^2) \right) \nonumber \\
\quad \quad &=& \int Tr (\xi * F) + i \int Tr \lbrack \upsilon \,, \xi *
F \rbrack_* + {\cal O}(\upsilon^2) \nonumber \\
\quad \quad &=& \int Tr (\xi * F) + i \int \upsilon^\alpha * \xi^\beta *
F^\gamma \, Tr \lbrack Q^\alpha \,, Q^\beta Q^\gamma \rbrack + i \int \,
Tr Q^\alpha Q^\beta Q^\gamma \, \lbrack \upsilon^\alpha \,, \xi^\beta * F^\gamma
\rbrack_* + {\cal O}(\upsilon^2) \nonumber \\   
\quad \quad &=& \int Tr (\xi * F) + {\cal O}(\upsilon^2) \,,
\eeqa
\label{eq:gaugeinv}
as was to be shown.

\vskip 1em

\noindent{\it Twisted Diffemorphisms:} \\

Implementation of space-time symmetries on noncommutative spaces was a 
long standing problem until very recently. It is well-known that on a
$d$-dimensional noncommutative space ${\mathbb R}_\theta^d$ generated
by the coordinates $x_\mu \in {\cal A}_\theta ({\mathbb R}^d)$, the 
Poincar\'e and diffeomorphism symmetries are explicitly broken due to 
the noncommutativity 
\be
\lbrack x_\mu \,, x_\nu \rbrack_* = i \theta_{\mu \nu} \, ,
\ee
if they are naively implemented.
Very recently it has been reported by Chaichian et. al.\cite{Chaichian}
and Aschieri et. al. \cite{wess} that these symmetries can be restored 
by twisting their coproduct (See also the earlier work of Oeckl \cite{Oeckl},
Such a twist in a general context is due to Drinfel'd \cite{Drinfeld}).
A clear way to understand these developments is as follows \cite{wess, Balachandran}. 

Let ${\cal A}$ be an algebra. ${\cal A}$ comes with a rule for multiplying
its elements. For $f, g \in {\cal A}$ there exists the multiplication map
$\mu$ such that 
\begin{gather}
\mu : {\cal A} \otimes {\cal A} \rightarrow {\cal A} \,, \nonumber \\
f \otimes g \rightarrow \mu (f \otimes g) \,.
\end{gather}

Now let ${\cal G}$ be the group of symmetries acting on ${\cal A}$ by a given
representation $D: g \rightarrow D(g)$ for $g \in {\cal G}$. We can denote this action by
\be
f \longrightarrow D(g) f \,.
\ee
The action of ${\cal G}$ on ${\cal A} \otimes {\cal A}$ is formally
implemented by the coproduct $\Delta(g)$. The action is compatible with $\mu$ 
only if a certain compatibility condition between $\Delta(g)$ and $\mu$ is 
satisfied. This action is 
\be
f \otimes g \longrightarrow  (D \otimes D) \Delta(g) f \otimes g \,,
\ee      
and the compatibility condition requires that 
\be
\mu \, \left ((D \otimes D) \Delta(g) f \otimes g \right ) = D(g) \, \mu  (f \otimes g) \,.
\label{eq:compatibility1}
\ee
The latter can be expressed neatly in terms of the following commutative diagram :
\begin{diagram}[width=6em]
f \otimes g & \rTo^{\Delta} & (D \otimes D) \Delta(g) f \otimes g\\
\dTo^{\mu} & & \dTo_{\mu} \\
\mu (f \otimes g) & \rTo^{} & D(g) \mu ( f \otimes g ) \\
\end{diagram}
If a $\Delta$ satisfying the above compatibility condition exists,
then ${\cal G}$ is an automorphism of ${\cal A}$. If such a $\Delta$ 
cannot be found, then ${\cal G}$ does not act on ${\cal A}$.   

We can now specialize to the algebra ${\cal A}_\theta ({\mathbb R}^d)$. 
The multiplication law on ${\cal A}_\theta ({\mathbb R}^d)$ is nothing 
but the Groenewold-Moyal $*$-product
\be
\mu_\theta (f \otimes g) = \mu_{\theta=0} \left({\cal F} \, f \otimes g \right) = f * g \,,
\ee
where $\mu_{\theta=0}(f \otimes g) = f g$ is the pointwise product and we
have introduced
\be
{\cal F} = e^{\frac{i}{2} \theta^{\mu \nu} \partial_\mu \otimes \partial_\nu} \,.
\ee 
The twisted coproduct is given by
\be
\Delta_\theta (g) = {\cal F}^{-1} \Delta_{\theta=0}(g) {\cal F} \,.
\ee 
$\Delta_\theta (g)$ satisfies the compatibility condition (\ref{eq:compatibility1}) 
with $\mu_\theta$ as can be easily checked.

For infinitesimal symmetries, it is sufficient to consider only the Lie
algebra $G$ of ${\cal G}$ and the coproduct associated with its universal
enveloping algebra.  

Infinitesimal diffeomorphisms are generated by vector fields of the
form $\zeta = \zeta^\mu(x) \partial_\mu$ and they form a Lie algebra.
The twisted coproduct of $\zeta$ is given by
\beqa
\Delta_\theta (\zeta) &=& {\cal F}^{-1} \Delta_0 (\zeta) {\cal F}
\nonumber \\   
&=& {\cal F}^{-1} (\zeta \otimes 1 + 1 \otimes \zeta) {\cal F} \,.
\eeqa
It is compatible with the multiplication map $\mu_\theta$ on ${\cal A}_\theta ({\mathbb R}^d)$. 

Variation of tensor fields under twisted diffeomorphisms can be
suitably represented by defining the operator $X^*_\zeta$ acting 
on ${\cal A}_\theta ({\mathbb R}^d)$ as
\be
{\hat \delta}_\zeta f  \equiv - X^*_\zeta (f) := - \zeta^\alpha \partial_\alpha f \,,
\label{eq:diff1} 
\ee
on scalars $f$. Its action on tensor fields follows by the standard 
transformation rules of tensors. For example, on a contravariant
vector field $V^\mu$ we have
\be
{\hat \delta}_\zeta V^\mu \equiv - X^*_\zeta (V^\mu) + 
X^*_{(\partial_\rho \zeta^\mu)} (V^\rho) := - \zeta^\rho
(\partial_\rho V^\mu) + (\partial_\rho \zeta^\mu) V^\rho \,.
\ee
We observe that the Leibniz rule for the ``twisted'' vector fields
is a deformed one. It is given by
\be
X^*_\zeta (f * g) = \mu_\theta \left \lbrace
{\cal F}^{-1} ( X^*_\zeta \otimes 1 + 1 \otimes X^*_\zeta) {\cal F}
\, (f \otimes g) \right \rbrace \,.
\ee
This deformation of the Leibniz rule ensures that the product of
two tensor fields of rank $n$ and $m$ transforms as a tensor field
of rank $n+m$. 

This much of information on ``twisted'' diffeomorphisms is sufficient
for our purposes. For further details on the subject we refer to \cite{wess}.

We are now ready to demonstrate the invariance of the action $S$ 
in (\ref{eq:action1}) under the "twisted" diffemorphisms. We have 
$\xi$ transforming as a scalar, and $\varepsilon^{\mu \nu} F_{\mu
  \nu}$ transforming as a tensor density of weight $-1$:
\begin{gather}
\delta_{{\hat \zeta}} \xi = - X_\zeta^* (\xi) := 
- \zeta^\alpha \partial_\alpha \xi \nonumber \\
\delta_{{\hat \zeta}} (\varepsilon^{\mu \nu} F_{\mu \nu}) = 
- X_\zeta^* (\varepsilon^{\mu \nu} F_{\mu \nu})
-X^*_{(\partial_\alpha \zeta^\alpha)} 
(\varepsilon^{\mu \nu} F_{\mu \nu}) \,.
\label{eq:transforms1}
\end{gather}
Using (\ref{eq:transforms1}) we find 
\be
\delta_{{\hat \zeta}} (\xi* \varepsilon^{\mu \nu} F_{\mu \nu}) =
- \partial_\alpha \big ( \zeta^\alpha (\xi * \varepsilon^{\mu \nu} 
F_{\mu \nu}) \big ) \,,
\label{eq:td1}
\ee    
or equivalently
\be
\delta_{{\hat \zeta}} (\xi * \varepsilon^{\mu \nu} F_{\mu \nu}) =
- \partial_\alpha \big ( X_{\zeta^\alpha}^* 
(\xi * \varepsilon^{\mu \nu} F_{\mu \nu}) \big )  \,.    
\label{eq:td2}
\ee
Thus under the infinitesimal ``twisted'' diffemorphisms generated 
by $\delta_{{\hat \zeta}}$, the Lagrangian changes by a total 
derivative:
\be
Tr (\xi * \varepsilon^{\mu \nu} F_{\mu \nu}) \longrightarrow 
Tr (\xi * \varepsilon^{\mu \nu} F_{\mu \nu}) - \partial_\alpha \big 
( X_{\zeta^\alpha}^* (Tr (\xi * \varepsilon^{\mu \nu} F_{\mu \nu})) \big )  \,, 
\ee
and hence the action $S$ is invariant.

\section{Classical Solutions}

\subsection{Equations of Motion}

Let us first examine the equations of motion following from the 
$S$ in (\ref{eq:action3}) when the fields $\eta_a, \eta_2, \eta_3$ 
are varied. We find 
\begin{subequations}\label{eq:nceoms}
\begin{align}
D^*e^a := d e^a + \frac{1}{2} {\varepsilon^a}_b (\omega \wedge_* e^b
- e^b \wedge_* \omega) & = 0 \,, \label{eq:i}
\\ 
d \omega - \frac{\Lambda}{2 s} \varepsilon_{ab} e^a \wedge_* e^b  & = 0 \,, \label{eq:ii}\\ 
d k + \frac{1}{2} \varepsilon_{ab} e^a \wedge_* e^b & = 0 \,.
\label{eq:iii} 
\end{align}
\end{subequations}

The commutative limit of these classical equations can immediately be 
obtained by replacing the $*$-product with the usual pointwise product. 
That gives the familiar equations
\begin{subequations}\label{eq:Commutativeeoms}
\begin{align}
D e^a = d e^a + {\varepsilon^a}_b \omega \wedge e^b & = 0 \,, \label{eq:iv}\\
d \omega - \frac{1}{2} \frac{\Lambda}{s} \varepsilon_{ab} e^a \wedge
e^b & = 0 \,, \label{eq:v} \\
d k + \frac{1}{2} \varepsilon_{ab} e^a \wedge e^b & = 0 \,.
\label{eq:vi}
\end{align}
\end{subequations}
We now study the solutions of the equations of motion (\ref{eq:nceoms})
for various values of the parameters $\Lambda$ and $s$.

\subsection{The $AdS_2$ solution} 

In this case we take both $\Lambda$ and $s$ to be finite and let
$\frac{\Lambda}{s} \rightarrow \Lambda $.

Let us first study the commutative theory from the equations of motion given in 
(\ref{eq:Commutativeeoms}). To this end first substitute $k = a - \frac{s}{\Lambda} \omega$ 
to (\ref{eq:vi}) and use (\ref{eq:v}) to find $da = 0 $. Thus the 1-form field $a$ is closed
and hence non-dynamical, and can be eliminated. Thus we set $a=0$ in what follows.
The equations of motion then describe standard two dimensional gravity
\cite{Isler, Cangemi1}. Making the substitution $k = a - \frac{s}{\Lambda} \omega$ in $A$
as well and letting $(J - \frac{1}{2}sI) \rightarrow J$, the gauge
field and the algebra can also be put into their conventional form. 
Thus the connection one form $A$ in (\ref{eq:connection}) can be rewritten as 
\be
A = e^a P_a + {\omega} J \,.
\label{eq:connection1form1}
\ee

For $\Lambda < 0$, this commutative model has the well-known $AdS_2$ solution
given by the metric
\be
ds^2 = \Lambda r^2 dt^2 - \frac{1}{\Lambda r^2} dr^2 \,.
\ee
It gives the connection 
\be
A_t = \frac{i \Lambda r}{2} 
\left (
\begin{array}{cc} 
1 & 1 \\
-1 & -1 \\
\end{array}
\right ) \,, \quad 
A_r = \frac{1}{2r}
\left (
\begin{array}{cc}
0 & 1 \\
1 & 0 \\
\end{array}
\right ) \,.
\
\label{eq:connection1form2}
\ee
The {\it zweibein's} and the spin connection can be read from  
(\ref{eq:connection1form2}) as 
\begin{gather}
e_t^0 = \sqrt{\Lambda} r \,, \quad e_t^1 = 0 \,, \quad e_r^0 =
0 \,, \quad e_r^1 = \frac{1}{\sqrt{\Lambda } r} \nonumber \\   
\omega_t = - \Lambda r \,, \quad \omega_r =0 \,.
\label{eq:components}
\end{gather}
It can be easily verified that (\ref{eq:components}) satisfies the equations 
of motion given in (\ref{eq:Commutativeeoms}). 

It has been shown in \cite{Klemm} that the above $AdS_2$ geometry is 
in fact a solution of the noncommutative version of the standard two dimensional
gravity model. Below we also reach this result by virtue of the presence of
twisted diffeomorphisms in our theory. 

Let us now show that (\ref{eq:components}) is indeed also a solution for
the NC equations of motion given in (\ref{eq:nceoms}). We reason that 
due to the invariance of our theory under twisted diffeomorphisms, the 
commutation relation between coordinates is preserved under
a general coordinate transformation. Thus we have
\be 
\lbrack x_0 \,, x_1 \rbrack = i \theta \quad 
\stackrel{\mbox{(twisted diffeos)}}{\longrightarrow}
 \quad \lbrack t \,, r \rbrack = i \theta \,.
\ee

We observe that the solution given in (\ref{eq:components}) is
time-independent. Hence all the $*$-products in (\ref{eq:nceoms}) 
collapse to pointwise products and these equations of motion are
satisfied immediately. This indeed shows that the $AdS_2$ geometry
is a solution to our noncommutative gauge theory.  

\subsection{The Black Hole Solution}

In this case we keep $\Lambda$ finite and take $s \rightarrow \infty$.
This is the NC version of the ``string inspired'' gravity model \cite{Cangemi1}.  

In order to discuss the solutions to the equations of motion in this case,
it is useful to express the action in terms of the light cone coordinates
\be
x^\pm = x^0 \pm x^1 \,.
\ee
Following Verlinde \cite{Verlinde}, this task can be carried out rather easily.
The action $S$ in (\ref{eq:action3}) takes the form
\begin{multline}
\int d x^+ d x^- \Big \lbrack \eta^+ * \big( d e^+ + \frac{1}{2} ( \omega \wedge_*
e^- - e^- \wedge_* \omega)\big) + \eta^- * \big ( d e^- - \frac{1}{2} 
(\omega \wedge_* e^+ - e^+ \wedge_* \omega) \big) + \\
\chi * d \omega + \eta_3 * d k + \eta_3 * e^+ \wedge_* e^- \Big \rbrack \,,
\label{eq:action4}
\end{multline}
where $\chi := \eta_2= e^\varphi$ and $\varphi$ is the dilaton field.

Let us first note that the variation of (\ref{eq:action4}) with respect to 
$k$ gives $\eta_3 = \mbox{constant}$. We set this constant 
equal to $\Lambda$. Variations with respect to the $e^-$ and $e^+$ give
\beqa
e^+ &=& -\frac{1}{\Lambda} D^* \eta^+ =  -\frac{1}{\Lambda} 
\big ( d \eta^+ - \frac{1}{2} ( \eta^+  * \omega + \omega * \eta^+))
\,, \nonumber \\
e^- &=&  \frac{1}{\Lambda} D^* \eta^- = \frac{1}{\Lambda} 
\big ( d \eta^- + \frac{1}{2} ( \eta^-  * \omega + \omega * \eta^-) \big ) 
\label{eq:e+-}
\eeqa
respectively, and from variation of $\omega$ we find  
\be
d \chi = - \frac{1}{2} \lbrace \eta^- , e^+ \rbrace_* +  \frac{1}{2}
\lbrace \eta^+ , e^- \rbrace_* \,.
\ee 
Let us also define
\be
M^* = -\Lambda \chi + \frac{1}{2} (\eta^+ * \eta^- +  \eta^- * \eta^+) \,.  
\label{eq:M}
\ee
In the commutative limit $M^*$ approaches the black hole mass $M$.
Using the above equations of motion we find  
\be
d M^* = \frac{1}{4} \big \lbrack \omega \,, \lbrack \eta^+ \,, \eta^-
\rbrack_* \big \rbrack_* + \frac{1}{4} \lbrack \eta^+ \,, \lbrack \omega \,, \eta^-
\rbrack_* \big \rbrack_* \,.
\label{eq:dM}
\ee

We recall that in the commutative theory the conformally scaled metric is 
given by \cite{Verlinde}
\be
{\widetilde G} = h_{ab} \frac{e^a \otimes e^b}{\chi} \equiv \frac{D \eta^+
\otimes D \eta^-} {-\frac{1}{\Lambda}(M - \eta^+ \eta^-)} \,. 
\label{eq:metric1}
\ee
We take the ansatz below as the natural generalization of (\ref{eq:metric1}) to the NC case:
\begin{multline}
{\widetilde G}^*_{\mu \nu } = \frac{1}{8}(D_\mu^*  \eta^+  * D_\nu^* \eta^- +
D_\nu^*  \eta^+  * D_\mu^* \eta^- ) * \left ( \frac{1}{- \frac{1}{\Lambda} 
\big ( M^* - \frac{1}{2} (\eta^+ * \eta^- +  \eta^- * \eta^+ )\big )} \right ) + \\
\frac{1}{8} \left ( \frac{1}{- \frac{1}{\Lambda} \big (M^* - \frac{1}{2} (\eta^+ * \eta^- + 
\eta^- * \eta^+ )\big )} \right ) * (D_\mu^*  \eta^+  * D_\nu^* \eta^- +
D_\nu^*  \eta^+ * D_\mu^* \eta^-) + \left( + \longleftrightarrow -\right) \,.
\label{eq:ncmetric}
\end{multline}
We note that ${\widetilde G}_{\mu \nu }$ as given above is symmetric and 
transforms as a second rank covariant tensor under ``twisted'' diffeomorphisms. 
Thus according to the definition given in \cite{wess}, it qualifies as a metric.
In what follows, we proceed by setting $\Lambda= -1$. 

In the commutative theory, the fields $\eta^\pm$ are related to light cone coordinates
$u$ and $v$ by
\be
u(x)= \eta^+(x) e^{- \int^x \omega} \,, \quad v(x)= \eta^-(x) e^{\int^x \omega} \,.       
\label{eq:uv}
\ee
In these coordinates the black hole metric and the dilaton are given by
\be
ds^2 = \frac{du dv}{1-uv} \,, \quad \varphi= \ln (1 - uv) \,,
\ee
where $M$ is set to be equal to $1$. $u$ and $v$ are also closely related to the Schwarzchild type of 
coordinates $t$ and $r$ by
\be
u = \sinh r e^t \,, v = -\sinh r e^{-t} \,.
\ee
The metric and the dilaton in these coordinates take the form
\be
ds^2 = dr^2 - \tanh^2 r dt^2 \,, \quad \varphi = \ln \cosh^2 r \,.
\label{eq:bhm}
\ee

Using the metric (\ref{eq:bhm}), it is easy to see that one has
\be
e_t^+ e_t^- = -\sinh^2 r \,, \quad e_r^+ e_r^- = \cosh^2 r  \,.
\ee

We see from (\ref{eq:uv}) that $\eta^+ \eta^- = -\sinh^2 r$ and thus is
time-independent. Let us make the gauge choice that $\eta^+$ and $\eta^-$ are time-independent
and only functions of $r$. The equations of motion of the commutative theory for
$e^-$ and $e^+$ then become   
\beqa
e^+_t = D_t \eta^+ &=& - \omega_t \eta^+ \nonumber \\
e^-_t = D_t \eta^- &=& - \omega_t \eta^- \,.
\eeqa
Thus
\be
e^+_t e^-_t = \eta^+ \eta^- \omega_t^2  \,,
\ee
and hence $\omega_t^2 = 1$ and 
\be
\omega_t = \pm 1 \,.
\label{eq:omegat}
\ee

The scalar curvature can be directly computed from the metric to be 
\be
R = \frac{4}{\cosh^2 r} \,.
\label{eq:scalarcur1}
\ee
It can also be expressed as
\be
R = \frac{2}{det~ e} \varepsilon^{\mu \nu} \partial_\mu \omega_\nu \,.
\label{eq:scalarcur}
\ee
Using this and (\ref{eq:omegat}) and (\ref{eq:scalarcur1}) in 
(\ref{eq:scalarcur}) and the result $det~ e = \tanh r$, we find 
\be
\partial_t \omega_r = - 2 \frac{\tanh r}{\cosh^2 r}
\ee
with the obvious solution
\be
\omega_r = - 2 \frac{\tanh r}{\cosh^2 r} t + h(r) \,.
\label{eq:omega}
\ee 

Once more we reason that due to the invariance of our theory under 
twisted diffeomorphisms, the canonical commutation relation between
coordinates are preserved under a general coordinate transformation.
Thus we have
\be 
\lbrack x_0 \,, x_1 \rbrack = i \theta \quad
\stackrel{\mbox{(twisted diffeos)}}{\longrightarrow} 
\quad \lbrack t \,, r \rbrack = i \theta \,.
\ee

Now using the above developments we find, expanding in powers of $\theta$:  
\beqa
D^* \eta^+ &=& d \eta^+ - \omega \eta^+ + \frac{1}{2} \theta^2 \varepsilon_{ij} 
\varepsilon_{kl} \partial_i \partial_k \omega \partial_j \partial_l \eta^+  
\nonumber + \mbox{higher order terms in}~\theta \\
&=& d \eta^+ - \omega \eta^+ = D \eta^+ \,.
\eeqa
The result above is exact to all orders in $\theta$ since odd powers in $\theta$ 
vanish anyway due to antisymmetry of $\theta_{\mu \nu}$ in its indices and even powers 
like the $\theta^2$ term above vanish after the differentiations on $\omega$.
Similarly we find $D^* \eta^- = D \eta^-$ to all orders in $\theta$. Time dependent 
solutions of the theory can be obtained through NC gauge transformations of $\eta^{\pm}$.
 
Thus we conclude that the black hole solution of the commutative theory, 
given by the metric in (\ref{eq:bhm}) is at the same time a solution
to our noncommutative theory. For this conclusion the presence of twisted 
diffeomorphisms is crucial as can be clearly observed from the arguments
above.
  
\section{Concluding Remarks}

In this article we have constructed noncommutative two dimensional gravities in
terms of noncommutative gauge theories. We have shown that the presence of
twisted diffemorphisms ensure that solutions of the corresponding commutative
gauge theories continue to be solutions to these noncommutative models. 
\vskip 1em

{\bf Acknowledgments} \\

A.P.B is supported in part by the DOE grant DE-FG02-85ER40231. 
T.R.G was on leave of absence from IMSc at DIAS and thanks Denjoe O'Connor
for making his visit possible and for the hospitality in DIAS, where 
part of this work was done. The work of S.K. is supported by Irish Research 
Council for Science Engineering and Technology (IRCSET). K.S.Gupta would like to 
thank S. K\"{u}rk\c{c}\"{u}o\v{g}lu and D. O'Connor for hospitality in DIAS where
a part of this work was done.

\end{document}